\documentclass[preprint,amsmath,amssymb,nofootinbib,superscriptaddress]{revtex4}

\usepackage{changes}
\usepackage{amsfonts}
\usepackage{amsmath,graphicx,color,epsfig}
\usepackage{CJK}

\newcommand{\be}{\begin{equation}}
\newcommand{\ee}{\end{equation}}
\newcommand{\bea}{\begin{eqnarray}}
\newcommand{\eea}{\end{eqnarray}}
\newcommand{\ba}{\begin{array}}
\newcommand{\ea}{\end{array}}
\newcommand{\bi}{\begin{itemize}}
\newcommand{\ei}{\end{itemize}}

\renewcommand{\vec}[1]{\mbox{\boldmath $#1 \!\!$ \unboldmath}}

\newcommand{\nslash}{\kern 0.2 em n\kern -0.50em /}
\newcommand{\kslash}{\kern 0.2 em k\kern -0.45em /}
\newcommand{\qslash}{\kern 0.2 em q\kern -0.45em /}
\newcommand{\pslash}{\kern 0.2 em p\kern -0.50em /}
\newcommand{\rslash}{\kern 0.2 em r\kern -0.50em /}
\newcommand{\sslash}{\kern 0.2 em s\kern -0.50em /}
\newcommand{\Sslash}{\kern 0.2 em S\kern -0.50em /}
\newcommand{\Pslash}{\kern 0.2 em P\kern -0.50em /}
\newcommand{\Dslash}{\kern 0.2 em D\kern -0.65em /\kern 0.15em}

\newcommand{\imp}{\affiliation{Institute of Modern Physics, Chinese Academy
of Sciences, Lanzhou 730000, China}}

\newcommand{\ucas}{\affiliation{University of
Chinese Academy of Sciences, Beijing 100049, China}}

\newcommand{\pucas}{\affiliation{School of Physical Sciences, University of
Chinese Academy of Sciences (UCAS), Beijing 100049, China}}

\begin{document}
\title{Identify the hidden charm pentaquark signal from non-resonant background
in electron-proton scattering}

\author{Zhi Yang\footnote{zhiyang@impcas.ac.cn}}
\imp \ucas

\author{Xu Cao\footnote{caoxu@impcas.ac.cn (Corresponding author)}}
\imp \ucas

\author{Yu-Tie Liang\footnote{liangyt@impcas.ac.cn}}
\imp \ucas

\author{Jia-Jun Wu\footnote{wujiajun@ucas.ac.cn}}
\pucas

\begin{abstract}

We study the electroproduction of the LHCb pentaquark states with the assumption that they are resonant states.
The main concern here is to investigate the final state distribution in the
phase space in order to extract the feeble pentaquark signal from the large
non-resonant background. Our results show that the ratio of the signal to
background would increase significantly with proper kinematic cut, which would be very helpful for future
experimental analysis.

\end{abstract}

\maketitle

\section{Introduction}

In the last decades, more and more possible candidates for exotic hadrons have
been experimentally established. Especially, in the year of 2015 the LHCb Collaboration announced the observation of two
pentaquark states, one narrow $P_c(4450)$ and one broad $P_c(4380)$, in the $J/\psi p$ invariant mass distribution in
the $\Lambda_b^0\to J/\psi K^- p$ decay~\cite{Aaij:2015tga}.
In 2019, the LHCb Collaboration updated the knowledge of the
pentaquarks with many more collected data samples of the same
decay~\cite{Aaij:2019vzc}.
A new narrow pentaquark candidate $P_c(4312)$ was
observed, while the old $P_c(4450)$ peak was found to be resolved into two
narrower structures, $P_c(4440)$ and $P_c(4457)$ owing to the larger statistics.
In a coupled-channel approach, it is argued the existence of a narrow
$P_c(4380)$ required by heavy quark spin symmetry~\cite{Du:2019pij}.
After their discovery, lots of discussion in association with their properties
have been triggered, and various interpretations were proposed for their
internal structure.
Since their masses are close to $\Sigma_c\bar{D}^{(*)}$ thresholds, many
literatures assigned them as $\Sigma_c\bar{D}^{(*)}$ molecular
states~\cite{Du:2019pij,Xiao:2019gjd,Xiao:2019mst,Meng:2019ilv,Burns:2019iih,Xu:2019zme,Wang:2019spc,Xu:2020gjl,Cheng:2019obk,Zhu:2019iwm,Zhang:2019xtu,Chen:2019asm,Liu:2019tjn,Chen:2019bip,He:2019ify}.
Alternative explanations include hadro-charmonium
states~\cite{Eides:2019tgv}, compact diquark-diquark-antiquark
states~\cite{Cheng:2019obk,Ali:2019npk,Ali:2019clg,Wang:2019got}, and etc.
A data driven analysis of the $P_c(4312)$ found it to be a virtual
state~\cite{Fernandez-Ramirez:2019koa}.
On the other hand, several recent works found that spin parity assignments for
$P_c$(4440) and $P_c$(4457) are sensitive to details of the one-pion exchange
potential~\cite{Liu:2019tjn,Yamaguchi:2019seo,Valderrama:2019chc,Du:2019pij}.
The fits of the measured $J/\psi p$ invariant mass distributions indeed point to different
quantum numbers for $P_c(4440)$ and $P_c(4457)$~\cite{Du:2019pij}.
Actually, the hidden charm pentaquark has been
predicted~\cite{Wu:2010jy,Wu:2010vk} before it was observed by LHCb.
Besides that, other possible pentaquarks in strange and bottom sector were also
suggested~\cite{Wu:2012wta,Shen:2017ayv,Xiao:2013jla,Karliner:2015ina}, though they are not yet observed
experimentally up to now. For more details, we refer to the comprehensive
reviews~\cite{Chen:2016qju,Guo:2017jvc,Lebed:2016hpi,Esposito:2016noz,Olsen:2017bmm,Liu:2019zoy,Brambilla:2019esw,Ali:2017jda,Guo:2019twa}.

However, since the discovery it was pointed out that the narrow peak of
pentaquark could be caused by the triangle
singularities~\cite{Guo:2015umn,Liu:2015fea} and it was further suggested
recently to distinguish them in isospin breaking decays~\cite{Guo:2019twa,Guo:2019fdo}.
Their decay and production properties are also extensively studied in various
scenarios ~\cite{Lin:2019qiv,Sakai:2019qph,Winney:2019edt,Chen:2020pac,Wu:2019adv,Wu:2019rog,Wang:2019krd,Guo:2019kdc}.
To discriminate their nature, the production of the pentaquark has
been proposed in
photo-induced~\cite{Wang:2015jsa,Huang:2016tcr,Blin:2016dlf,Karliner:2015voa,Kubarovsky:2015aaa}
and pion-induced
reactions~\cite{Lu:2015fva,Liu:2016dli,Kim:2016cxr,Wang:2019dsi}, because the
triangle singularity can not be present in two-body final states of the
production process. Thus if they are observed in the $J/\psi p$ or
open charm production~\cite{Huang:2016tcr}, they should be genuine states other than
kinematic effects. Later, the experimental search of pentaquark through
photoproduction was proposed at JLab~\cite{Meziani:2016lhg}. The GlueX
Collaboration searched for the pentaquark states through the near-threshold
$J/\psi$ exclusive photoproduction off the proton~\cite{Ali:2019lzf}. No
evidence for pentaquark photoproduction was found, and the model-dependent upper
limits on their branching fraction $\mathcal{B}(P_c\to J/\psi p)$ was set. The
photoproduction rate was investigated in model-dependent
calculations~\cite{Cao:2019kst,Wu:2019adv}, where the coupling of $P_c$
radiative decay was evaluated by the vector meson dominance (VMD) model. Though the 
extracted branching ratio $\mathcal{B}(P_c\to J/\psi p)$ is dependent on the
details of the VMD, e.g. the off-shell form factor, the photoproduction rate tends to be 
not large compared to the non-resonant contribution.
The double polarization observables were proposed to be useful in the search of pentaquark
photoproduction~\cite{Winney:2019edt}. However, the LHCb results indicate a
model-independent lower limit of $\mathcal{B}(P_c\to
J/\psi p)$~\cite{Cao:2019kst}, so it is hopeful to find the pentaquark eletro- and 
photo-production after enough events are accumulated if $P_c$ is real resonant state.
Moreover, it is expected that the distributions of $J/\psi$ for 
pentaquark and pomeron are different in the large angles of differential cross
sections~\cite{Wang:2015jsa,Wu:2019adv}, which could be helpful to identify the
pentaquark in cross sections.

After the update of JLab accelerator to 12 GeV, the search for pentaquark electroproduction at JLab12 will continue.
Recently an electron-ion collider at China (EicC) is proposed and hadron physics
is one of its main concerns~\cite{Cao:2020EicC}. Its designed center of mass
(c.m.) energy 15 $\sim$ 20 GeV covers the charmonium electroproduction.
In this paper, we will investigate the electroproduction of the the pentaquark
states at these machines. The main concern here is on the final state distributions, from which
a kinematic cut would isolate the feeble pentaquark signal from the large
non-resonant background. This paper is organised as follows. In Sec.~\ref{sec:anaframe}, we
will briefly describe the analytic formalism in our computation, following which
the results and discussions are given in Sec.~\ref{sec:numresults}. At last, we
will present a short summary.

\section{Formalism}
\label{sec:anaframe}

\begin{figure}[tb]
\begin{center}
  \includegraphics[width=0.8\textwidth]{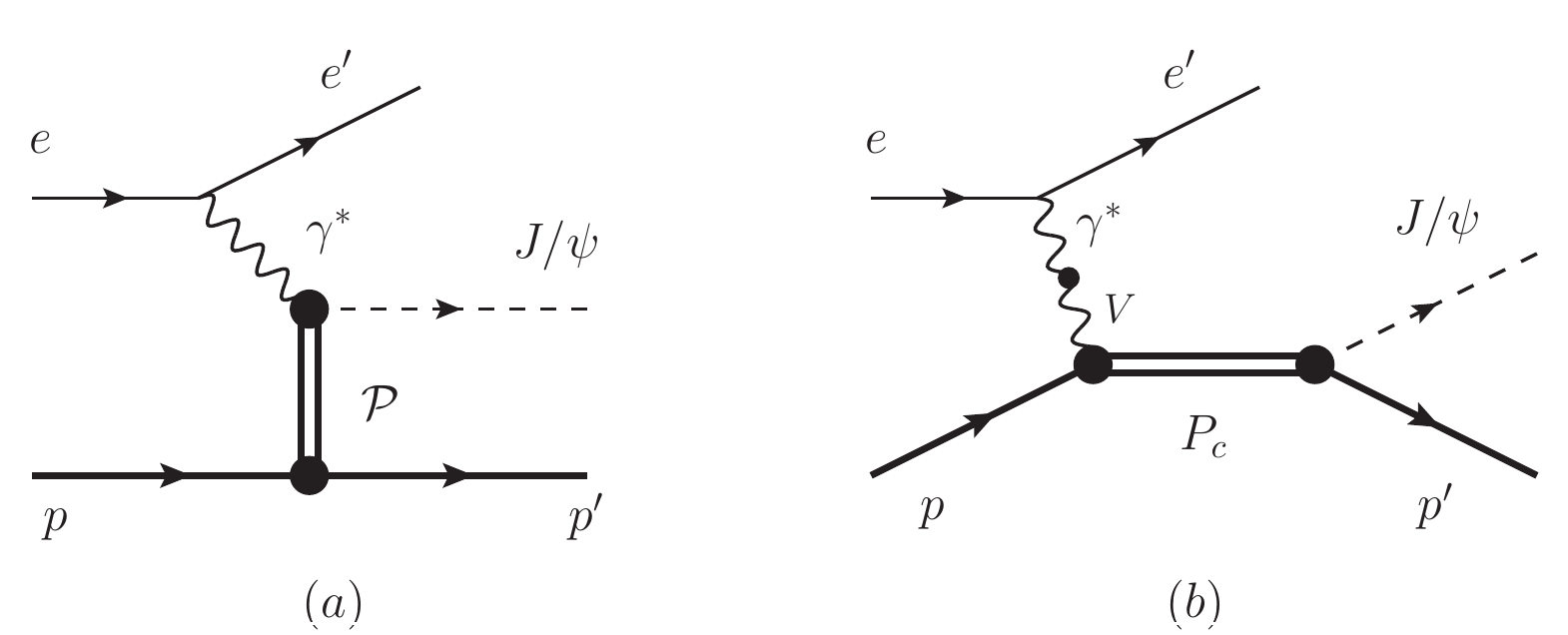}
\caption{Diagrams for the electroproduction of heavy quarkonium $J/\psi$. (a): The contribution of $t$-channel Pomeron exchange.
(b): pentaquark $P_c$ production in $s$-channel, where $V$ stands for the all possible
vector mesons.}
\label{fig:Diagram}
\end{center}
\end{figure}

As shown in Fig.~\ref{fig:Diagram}, the $pJ/\psi$ in final states of $e p \to e
p J/\psi$ can be produced from pentaquark decay (b) and the non-resonant
$t$-channel (a). Here the $u$-channel contribution is from $P_c$ or
$p$ exchange, but both of them are negligible since the highly off-shell
intermediate $P_c$ state and very small coupling between $J/\psi p p$,
respectively. Several phenomenological models were constructed to parameterize the $t$-channel diagram with gluon or Pomeron exchange. A detailed
comparison of these models can be found in Ref.~\cite{Cao:2019gqo} for the
$\Upsilon$ photoproduction. Here we take the soft dipole Pomeron model, which
can describe vector meson photoproduction from low to high
energies~\cite{Martynov:2002ez}.
We use a covariant orbital-spin (L-S) scheme to construct Langrangians of $P_c$
couplings~\cite{Zou:2002yy}, which has been used widely for the normal $N^*$ and
$\Delta^*$ resonances~\cite{Cao:2010ji,Cao:2010jj,Cao:2010km}.

\subsection{Pomeron exchange}

The Pomeron exchange model~\cite{Donnachie:1987pu,Pichowsky:1996jx,Laget:1994ba}
accounts for the dominant contribution in the leptoproduction process. The
Pomeron mediates the long range interaction between the nucleon and the confined
(anti-)quarks within the quarkonium. This is an effective and useful model to
parameterize the diffractive process for the production of neutral vector
mesons in the high energy region. By including a double Regge pole with intercept equal to one,
the soft dipole Pomeron model does not violate unitarity bounds and
can describe nearly all available cross-sections data of photo-
and electro-production of vector meson from light to heavy and from near threshold to high energies region in a consistent manner~\cite{Martynov:2002ez}.

\begin{figure}[tb]
\begin{center}
  \includegraphics[width=0.7\textwidth]{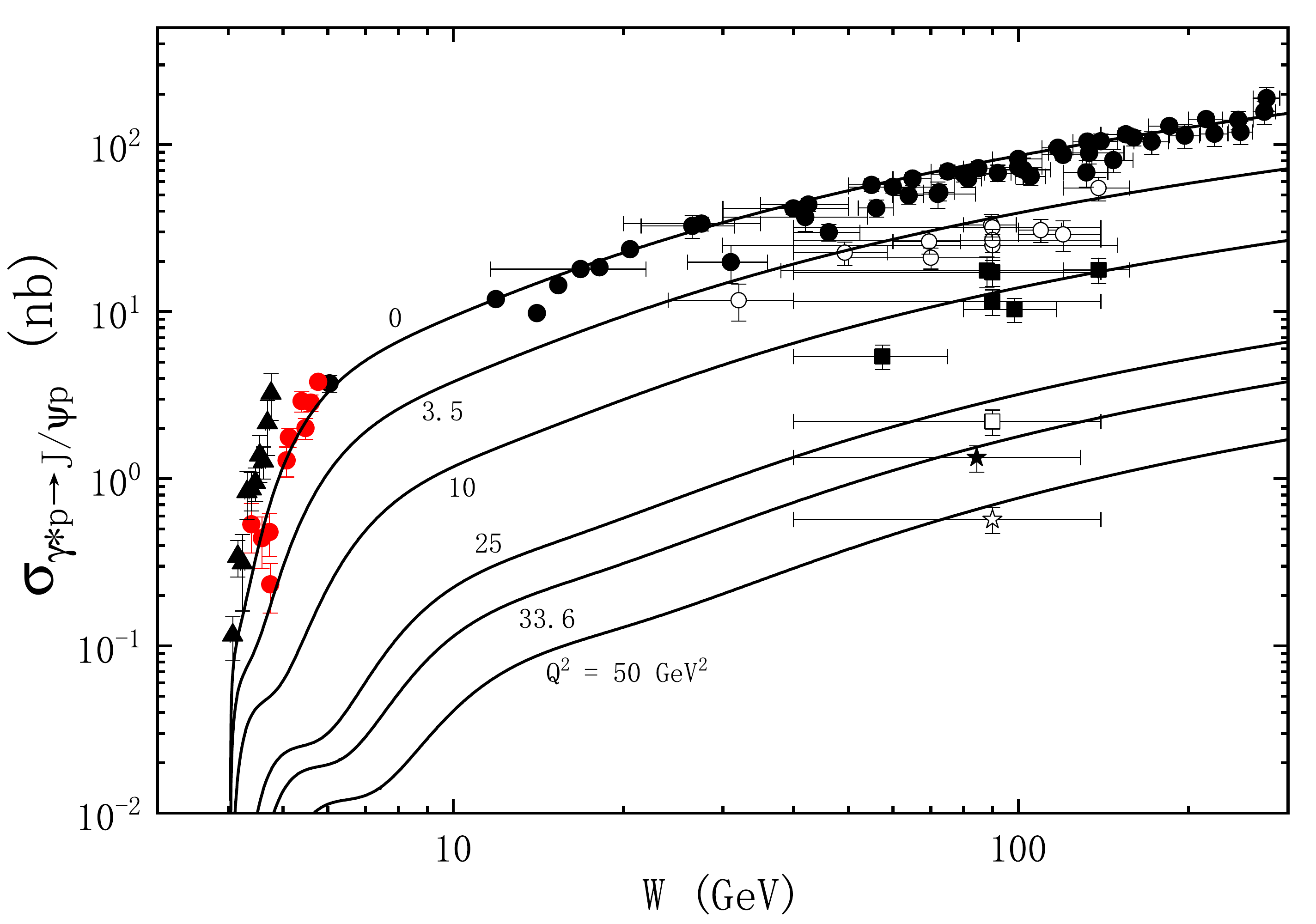}
\caption{The photonproduction cross section in terms of the invariant mass of
$J/\psi$ proton system. The near-threshold data with error bar
is the experimental measurements from GlueX~\cite{Ali:2019lzf} (solid triangle),
SLAC~\cite{Camerini:1975cy} (red circle), while the other data is from
H1~\cite{Aid:1996dn,Adloff:1999zs,Adloff:2000vm} and
ZEUS~\cite{Breitweg:1998nh,Chekanov:2002xi}.
The line is the fit from dipole Pomeron.}
\label{fig:pomeron}
\end{center}
\end{figure}

We start from the photoproduction of vector meson $V$ off proton in the soft dipole
Pomeron model with the formula of $t$-dependence cross section~\cite{Martynov:2002ez}
\be
\frac{d\sigma}{dt}=4\pi\,|\mathcal{M}^\mathcal{P}_{\gamma p\to Vp}|^2,
\label{eq:dipomeron}
\ee
where the amplitudes are defined as
\bea
\mathcal{M}^\mathcal{P}_{\gamma p\to Vp} &=& \mathcal{P}(z,t,M_V^2,Q^2) +
\mathcal{F}(z,t,M_V^2,Q^2),
\\
\mathcal{P}(z,t,M_V^2,Q^2) &=& i g_0 (-iz)^{\alpha_P(t)-1} + i g_1
\textrm{ln}(-iz) (-iz)^{\alpha_P(t)-1}, \\
\mathcal{F}(z,t,M_V^2,Q^2) &=& i g_f (-iz)^{\alpha_f(t)-1}.
\eea
The $\mathcal{P}$ and $\mathcal{F}$ terms are the so called dipole Pomeron and
Reggeon. The $Q^2 = -q^2$ and $M_V$ are photon virtuality and mass of vector meson, respectively.
The variable $z \sim \cos \theta$ with $\theta$ being the scattering angle of
final states in c.m. system of $\gamma^* p$. The nonlinear Pomeron
trajectory is $\alpha_{\mathbb{P}}(t)=1+\gamma(\sqrt{4 m_{\pi}^{2}}-\sqrt{4
m_{\pi}^{2}-t})$ with $m_{\pi}$ the pion mass and the Reggeon trajectory is
$\alpha_f(t)=\alpha_{f}(0)+\alpha^{\prime}_{f}(0)t$ with $\alpha_{f}(0)=0.8$
and $\alpha^{\prime}_{f}(0)=0.85\;\text{GeV}^{-2}$.
The parameters $\gamma=0.05\;\text{GeV}^{-1}$, $g_0=-0.03$, $g_1=0.01$ and
$g_f=0.08$ can be obtained by fitting the vector meson photoproduction. More
details can be found in Ref.~\cite{Martynov:2002ez}.
The advantage of this model is that it includes exclusive photoproduction of all
vector mesons for both real and virtual photons, as shown in above amplitudes.
This is convenient for our calculation of electroproduction here. The results
for the $J/\psi$ photoproduction are shown in Fig.~\ref{fig:pomeron}.

The electroproduction amplitude for the Pomeron exchange is evaluated as
\bea
\mathcal{M}_{ep\to eVp} &=&
M_{R_1}^{\mu}\frac{-g_{\mu\nu}}{q^2}\mathcal{M}_{R_2}^{\nu},
\label{eq:lepto}
\eea
where $R_1$ is the sub-reaction $e\to e\gamma$ and $M_{R_1}^{\mu} =i e
\bar{u}(k') \gamma^\mu u(k)$, while $R_2$ is
the sub-reaction $\gamma p\to Vp$.
If neglecting polarization correlations between the two sub-reactions, we have
the amplitude square
\bea
|\mathcal{M}_{ep\to eVp}|^2 &=& \frac{1}{3(q^2)^2}
\sum_{\lambda_1,\lambda_2}|M_{R_1}^{\mu}\epsilon_{\mu}^{*\lambda_1}|^2
|\mathcal{M}_{R_2}^{\nu}\epsilon_{\nu}^{\lambda_2}|^2,
\eea
where $\epsilon$ is the polarization vector of intermediate photon with spin of
$z$-direction $\lambda_{1,2}$.
The amplitude for the sub-reaction $R_2$ can be determined from the differential
cross section $d\sigma/dt$ by dipole Pomeron model mentioned in
Eq.~\eqref{eq:dipomeron} with the relation $|\mathcal{M}^\mathcal{P}_{\gamma p\to
Vp}|^2=\sum_{\lambda_2}|\mathcal{M}_{R_2}^{\nu}\epsilon_{\nu}^{\lambda_2}|^2$.
Alternative approach to investigate the electroproduction of vector meson is
using a microscopic description of the
Pomeron exchange~\cite{Donnachie:1984xq,Kim:2020wrd}.

\subsection{Pentaquark}

Here we only consider the $P_c$ with quantum
numbers $\frac{3}{2}^-$ in line with the GlueX~\cite{Ali:2019lzf},
where the branching fractions 
were determined by using the JPAC model~\cite{Blin:2016dlf}. Similar conclusion would be driven for alternative assignment of $J^P$.
The effective Lagrangian for the coupling of $P_c$ to $J/\psi p$ is written
as~\cite{Cao:2010ji,Wu:2019adv}
\bea
{\cal L}^{3/2^-}_{V BR} &=& g \overline{B} \vec\tau
\cdot \vec{V^{\mu}} R_{\mu} + h.c., \label{eq:VNP13}
\eea
where $R$ and $B$ denote $P_c$ resonance and the nucleon, respectively. The
coupling constant $g$ can be determined from the corresponding decay widths.
Here we use the total decay widths of $P_c$ as the measured values by LHCb and the upper
limits of branching fractions $\mathcal{B}(P_c\to J/\psi p)$ determined by
GlueX~\cite{Ali:2019lzf}. The propagator of $P_c$ can be written as
\bea
G^{3/2}_{R}(p_R)&=&\frac{ -i (\not\! p_R + M_R)
G_{\mu\nu}(p_R)}{p_R^2-M^2_{R}+iM_{R}\Gamma_{R}},
\eea
where $p_R$ is the momentum of the propagator, and $M_R$ the mass,
$\Gamma_{R}$ the decay width of $P_c$. The term $G_{\mu\nu}(p_R)$ is defined as
\bea
G_{\mu \nu}(p_R) &=& - g_{\mu \nu} + \frac{1}{3} \gamma_\mu \gamma_\nu +
\frac{1}{3 M_R}( \gamma_\mu p_{R\nu} - \gamma_\nu p_{R\mu}) + \frac{2}{3
p_R^2} p_{R\mu} p_{R\nu}.
\eea

We assume that the pentaquark resonances couple to photon via the vector meson pole
by using VMD model. Therefore the $\gamma p\to P_c$ vertex can be considered as $\gamma
p\to V p\to P_c$ as shown in Fig~\ref{fig:Diagram}(b). The coupling of vector
meson to photon is
\begin{eqnarray}
\mathcal{L}_{V\gamma}=\sum_V \frac{e M_V^2}{f_V} V_\mu A^\mu,
\label{eq:vmd}
\end{eqnarray}
where $M_V$ is the mass of vector meson, and $V^{\mu}$ and $A^\mu$ are the
vector meson and photon field, respectively.
Then the coupling constant of vector meson to photon $e/f_V$ can be extracted from
the partial decay width $\Gamma_{V\to e^+e^-}$ from the formula
\begin{equation}
 \frac{e}{f_V} =
\left[\frac{3 \Gamma_{V\to e^+e^-}}{2 \alpha_{em} |{\bf
p}_e|}\right]^{\frac{1}{2}},
\end{equation}
where the mass of electron and positron have been neglected, and ${\bf p}_e$ is the
three-vector momentum of electron in the vector meson rest frame.

For the off-shell vector meson in VMD, we choose the form factor
\be
\mathcal{F}(q^2)=\frac{\Lambda^4}{\Lambda^4+(q^2-M_V^2)^2},
\ee
where $\Lambda$ is the cut-off parameter. The choice of the vector
meson and the cut-off parameter would not change the distribution of the final
state, which is the main concern here.
Thus we choose the vector meson to be $J/\psi$ and the cutoff $\Lambda=0.5$ GeV.

\section{Results and discussions}
\label{sec:numresults}

\begin{table}
\centering
 \begin{tabular}{c|c|c|c}
\hline\hline
   & $M$ [MeV] & $\Gamma$ [MeV] & $J^P$   \\
\hline
$P_c$(4312) & $4311.9\pm0.7^{+6.8}_{-0.6}$ & $9.8\pm2.7^{+3.7}_{-4.5}$
& $\frac{3}{2}^-$
\\
$P_c$(4440) & $4440.3\pm1.3^{+4.1}_{-4.7}$ & $20.6\pm4.9^{+8.7}_{-10.1}$
& $\frac{3}{2}^-$
\\
$P_c$(4457) & $4457.3\pm0.6^{+4.1}_{-1.7}$ & $6.4\pm2.0^{+5.7}_{-1.9}$
& $\frac{3}{2}^-$
\\
\hline\hline
\end{tabular}
\caption{The measured masses and widths by LHCb~\cite{Aaij:2019vzc}, while the
quantum numbers are taken to be in line with the GlueX~\cite{Ali:2019lzf}
since the branching fractions were used here.}
\label{tab:Pcs}
\end{table}

We explore the electroproduction of the pentaquarks
observed by LHCb Collaboration as listed in Tab.~\ref{tab:Pcs}, together with
the contributions from Pomeron exchange, at JLab12 and EicC energy configurations.
JLab12 is the fixed-target experiment with $12$ GeV electron and rest proton,
while EicC is the colliding one with $3.5$ GeV electron and $20$ GeV proton.
The pentaquark and Pomeron contributions were added incoherently.
The interference terms may have large
contribution on the total cross section of pentaquark, but distribute smoothly in phase space. 
Also it is too premature to consider the interference at present, 
because we do not know the relative phase between different contributions.
Most importantly, these terms can be neglected for searching pentaquark 
since we will focus on the pentaquark dominant phase space area, from which we
will get the main conclusion in this paper. In our calculation, we choose the laboratory frame
with the electron moving in opposite $z$ direction. The cross sections were evaluated by using the VEGAS
program~\cite{Lepage:1977sw} which numerically integrates the kinematic events
generated by RAMBO~\cite{Kleiss:1985gy} with the dynamics described by the formula
above. Besides that, we get the final state distributions at the same
time.

The total production cross sections for both JLab12 and EicC are shown in
Tab.~\ref{tab:Xs}. We can see that the cross sections of non-resonant background
are a few orders of magnitude larger than that of pentaquarks.
For the cross section of pentaquarks, the model-dependent branching fractons
determined by GlueX and the cut-off parameter in the form factor appear as
overall factor, which means the cross sectons rather than the final state
distributions are greatly model-dependent. Thus the distributions are our main
concern here.

\begin{table}
\caption{The electroproduction cross section (in unit of pb) of pentaquarks and
non-resonant backgound at JLab12 and EicC.}
\centering
\begin{tabular}{|c|cc|}
  \hline\hline
   & Background & Pentaquarks  \\\hline 
  JLab12  & 1.4 & 0.0016    \\ 
  EicC    & 111   & 0.013    \\
  \hline\hline
\end{tabular}
\label{tab:Xs}
\end{table}

\begin{figure}[tb]
\begin{center}
  \includegraphics[width=0.95\textwidth]{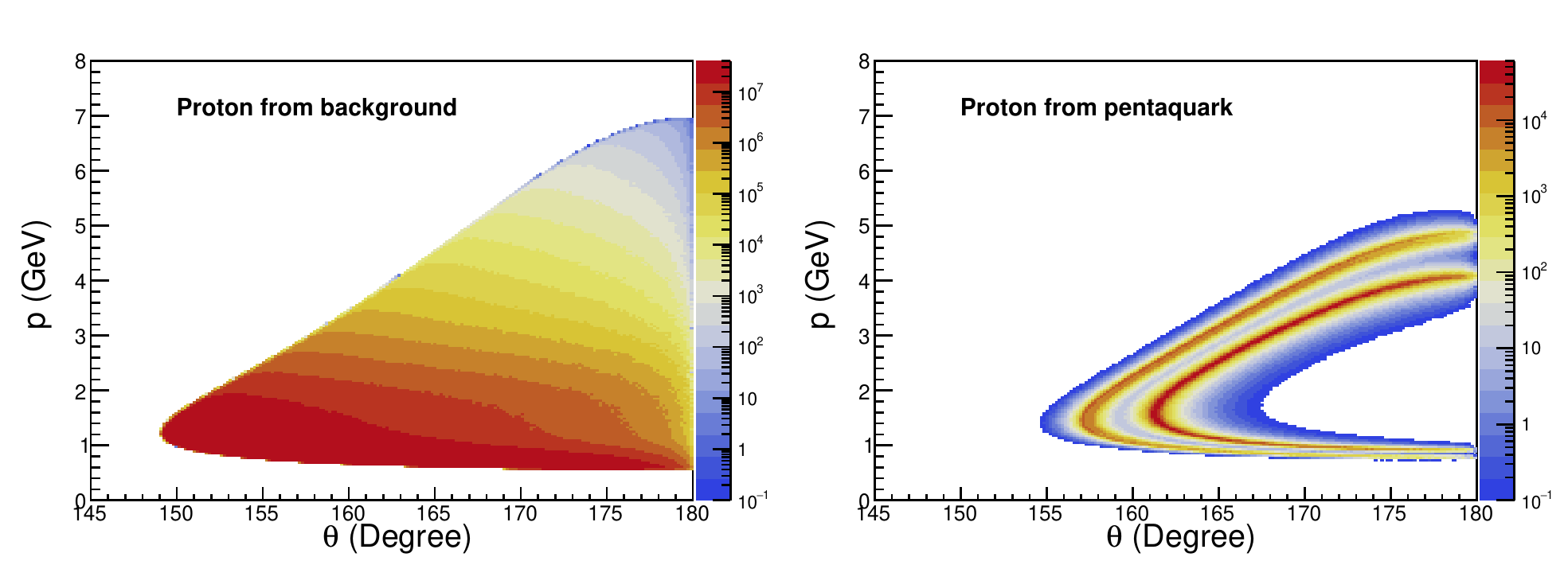}
\caption{The distribution of momentum versus scattering angles of final proton
in the laboratory frame at the JLab energy configuration, with 12 GeV electron
beam projectile on rest proton. We set the electron beam moving in opposite $z$
direction. The left panel is the final proton from Pomeron exchange, while the
right one is from pentaquark. The colors represent the differential cross
section.}
\label{fig:phasespace_jlab}
\end{center}
\end{figure}

\begin{figure}[tb]
\begin{center}
  \includegraphics[width=0.95\textwidth]{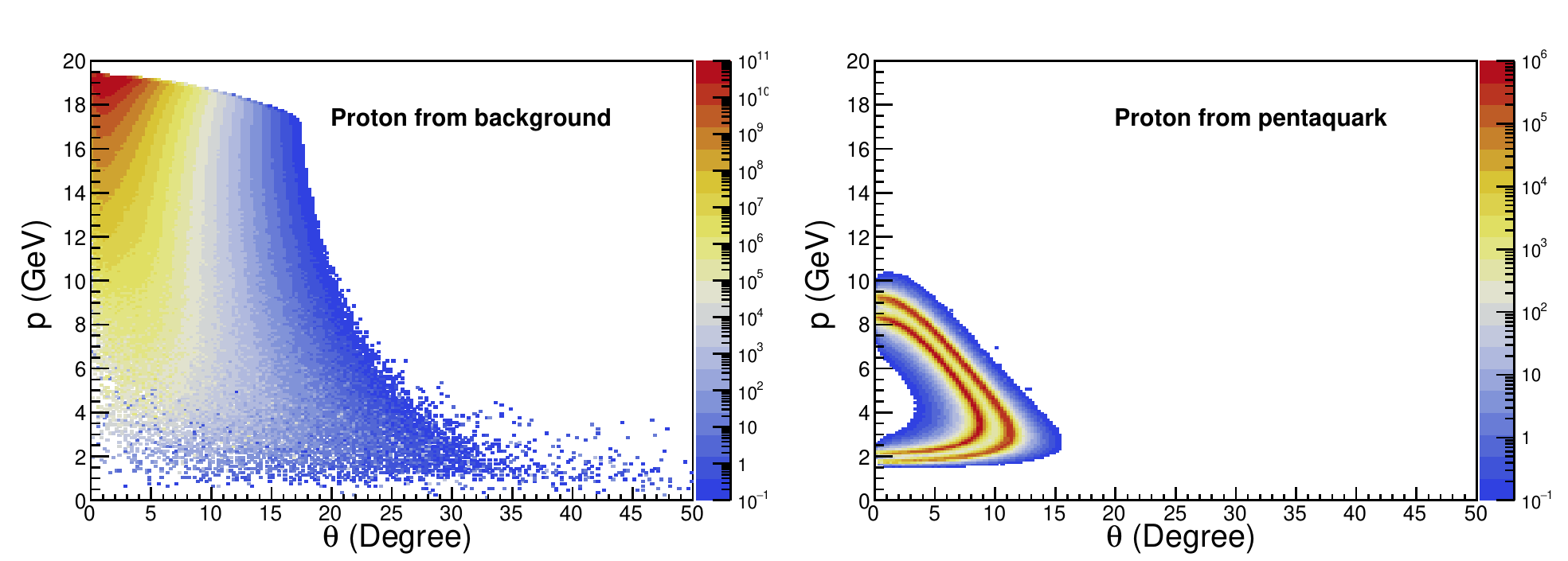}
\caption{Same as Fig.~\ref{fig:phasespace_jlab} but at EicC energy
configuration, which is 20 GeV proton colliding with 3.5 GeV electron. For a
better comparison, we choose the same range in axis of the two panels.}
\label{fig:phasespace_eicc}
\end{center}
\end{figure}

The three-momentum
and polar angle distributions of the final proton from either Pomeron 
exchange process or pentaquark production are shown separately in
Fig.~\ref{fig:phasespace_jlab} at JLab12 case and in
Fig.~\ref{fig:phasespace_eicc} at EicC, respectively. The left panels are for
the final proton from Pomeron contribution, while the right ones are for the
proton from the pentaquark decay. For a better comparison, we use the same
range in axes of the two panels for each figure. Due to the totally different
energy configuration, the final proton moves in the electron and proton forward angle
at JLab12 and EicC, respectively. As we can see, the polar angle distributions of final proton are very different for Pomeron exchange and pentaquark production. This is because the $t$-dependent cross section in
Eq.~(\ref{eq:dipomeron}) is suppressed at large $t$ for Pomeron exchange, while
the shape from pentaquarks is all flat across the full $t$ range.  This fact has
been already pointed out by several
papers~\cite{Wang:2015jsa,Cao:2019kst,Wu:2019adv} that the final particles from
different contributions have different behaviour at large angles.

In both energy configurations of JLab12 and EicC, the distribution of proton
decaying from pentaquarks are quite similar in shape but different in range.
The three pentaquarks are characteristic by the obvious resonant bands.
Among them, the $P_c$(4457) and $P_c$(4440) overlap with each other
because of their closeness of masses, so a quite good
energy resolution is needed to distinguish them. This is a challenge for future
detector design.

\begin{figure}[b]
\begin{center}
  \includegraphics[width=0.5\textwidth]{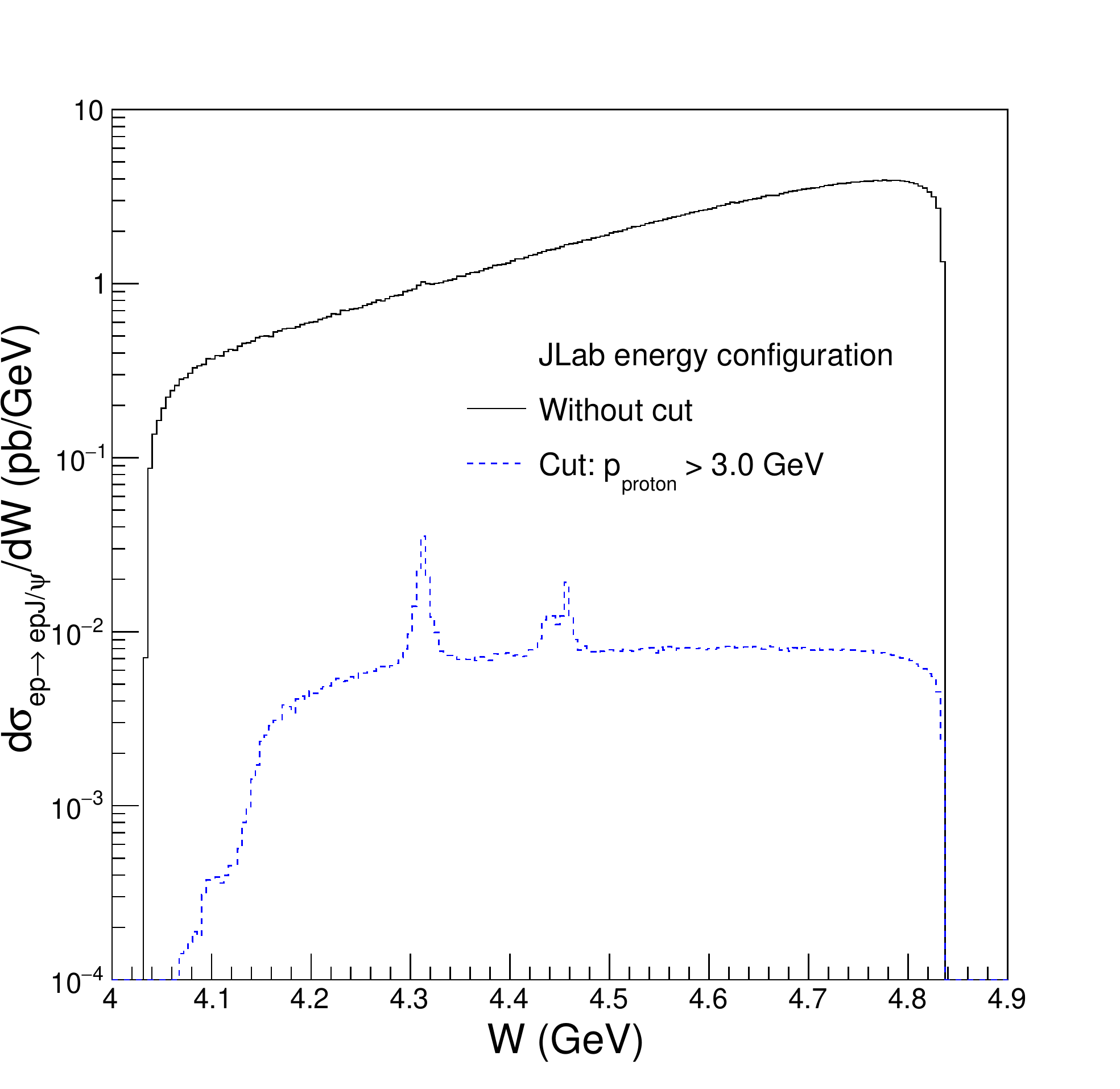}\hfill
  \includegraphics[width=0.5\textwidth]{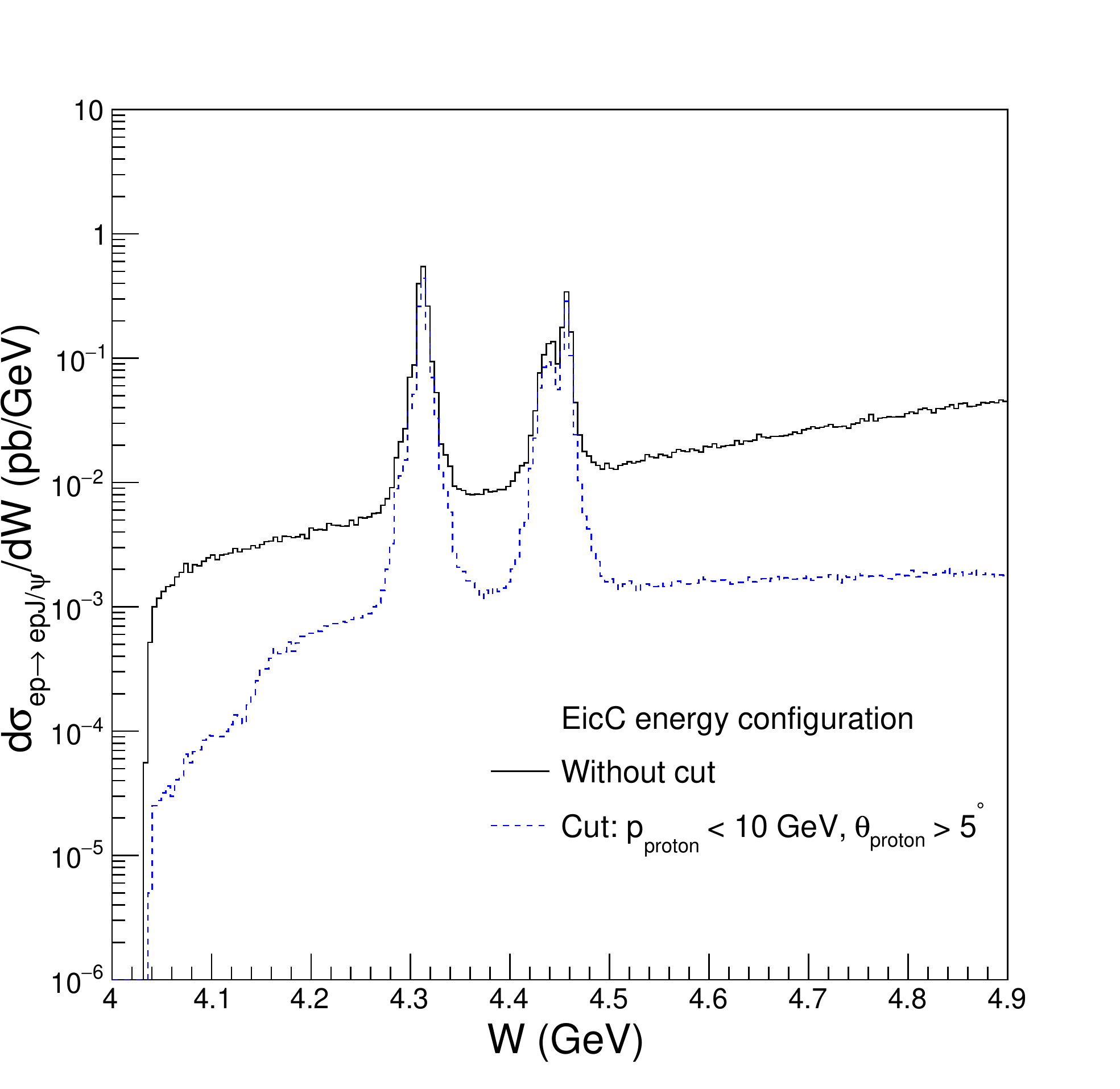}
\caption{The differential cross section of the electroproduction process in
terms of the invariant mass of $J/\psi$ proton system at
JLab (left) and EicC (right), respectively. The solid and dashed lines are for
the one without and with the cut, respectively. The peaks correspond to the three pentaquarks.}
\label{fig:diff}
\end{center}
\end{figure}

Most importantly, for each energy configuration, the proton from non-resonance
background and pentaquark present significant differences in phase space shown
in Fig.~\ref{fig:phasespace_jlab} and Fig.~\ref{fig:phasespace_eicc}.
As a result, we can take advantage of this feature to enhance the $P_c$ peaks
relative to that of the Pomeron exchange. It comes along with the main
conclusion of this paper in Fig.~\ref{fig:diff}, which shows the differential
cross section of the electroproduction process at the energy configuration of
JLab12 and EicC. The dashed and solid curves in Fig.~\ref{fig:diff} show the
results with and without the cut on the three-momentum and angle of final
proton, respectively. As we can see, a simple cut $p > 3\;\text{GeV}$ can remove
much more Pomeron contribution than pentaquark for JLab12. While for EicC, the
cut $p <10\;\text{GeV}$ and $\theta>5^{\circ}$ also works very well to depress the background.
Quantitatively, in the case of $P_c(4312)$ at JLab12, a signal over background
ratio increases from 0.3 to 19 with the kinematic cut. Therefore the
kinematic cut can make the $P_c$ peaks obviously more prominent and present
the huge potential in experimental analysis although the total number
of events would decrease after the cuts are used. More complex cuts would make
the situation much better.

At last we want to emphasize the great potential of both EicC and
JLab12 to search for the pentaquark. EicC has a higher signal over background
ratio, while JLab12 has much higher luminosity. The center of mass energy of EicC is about 16.7 GeV, much
larger than 4.8 GeV of JLab12.
As shown in Tab.~\ref{tab:Xs}, the larger center of mass energy would make the
total cross section 8 and 80 times larger for pentaquark signal and
non-resonant background, respectively. However EicC has 15 times larger
phase space in the invariant mass $W$ for the background than JLab.
Therefore the pentaquark signal could be presented more prominently in the
differential cross section at EicC in Fig.~\ref{fig:diff}.
On the other hand, the differential cross section is less reduced at EicC than
that at JLab12 after the kinematic cut is employed, which shows that colliding
mode could be better to study pentaquark than fixed-target mode.

\section{Summary}
\label{sec:summary}
\bigskip

The GlueX Collaboration at JLab has searched for the
pentaquark photoproduction and got negative result at present precision~\cite{Ali:2019lzf}.
One possibility is that the pentaquark
signal is small in total cross section compared to the non-resonant
contribution.
The production rates have been already investigated in many efforts~\cite{Cao:2019kst,Wu:2019adv}
and the signal of pentaquark in hidden charm photoproduction would be really small in cross sections.

In this paper,
we calculated the $ep\to e^{\prime}p^{\prime}J/\psi$ process with both
non-resonant $t$-channel contribution and hidden charm pentaquark in $s$-channel.
After the non-resonant
contribution is normalized using the soft dipole Pomeron model by photoproduction
data, the distribution of final particles from both sources were investigated.
In view of the very different shape of final proton in phase space owing to the
different underlying mechanism, we proposed that the three-momentum and angle
cuts on proton could largely suppress the non-resonant contribution and the
signal/background ratio would be significantly increased. For both energy
configuration of JLab12 and EicC, we found promising strategies even with
simple cuts, which is very helpful for future experimental analysis of the
electroproduction at these machines. In addition, it is also
promising to search for pentaquark at higher energy collider, e.g. US EIC.
Similar cut like the one used for EicC would also be helpful. Our criterion is
also enlightening for the electroproduction of $P_b$, bottom analog of $P_c$
states~\cite{Cao:2019gqo}.

Here we focus on the method to depress the background rather than the total
production cross section of pentaquarks, because the total cross section of
pentaqurks is greatly model-dependent due to the unknown coupling constant and
cut-off parameter appearing as overall factors.
Last but not least, we shall point out that our
framework could be used in the full simulation, the selection criterion of final
particles and optimization of detector design in the future.

\section*{Acknowledgments}

Z. Yang gratefully acknowledge the hospitality at the ITP where part of this
work was performed. We are grateful to F. K. Guo, Q. Wang, Q. Zhao and B. S. Zou
for useful discussions and comments. This work was supported in part by the National Natural Science
Foundation of China (Grant Nos. 11405222, 11975278), by the the Pioneer Hundred
Talents Program of Chinese Academy of Sciences (CAS) and by the Key Research
Program of CAS (Grant No. XDPB09).


\end{document}